\documentclass[aps,prl,twocolumn,floatfix,preprintnumbers,nofootinbib, superscriptaddress]{revtex4-1}

\usepackage{blindtext}
\usepackage{epsf,graphicx,xcolor}
\usepackage{amsmath,amssymb,amstext}
\usepackage{slashed}
\allowdisplaybreaks
\usepackage{verbatim}

\usepackage{mathtools}
\usepackage{epstopdf}
\usepackage{url,hyperref}
\usepackage[utf8]{inputenc}
\usepackage[english]{babel}
\usepackage{float}
\usepackage[compat=1.1.0]{tikz-feynman}

\begin{document}

\title{X17 discovery potential in the $\gamma N \rightarrow e^+ e^- N$ process at electron scattering facilities}

\author{Johannes Backens}
\affiliation{Institut f\"ur Kernphysik and $\text{PRISMA}^+$ Cluster of Excellence, Johannes Gutenberg-Universit\"at, D-55099 Mainz, Germany}
\author{Marc Vanderhaeghen}
\affiliation{Institut f\"ur Kernphysik and $\text{PRISMA}^+$ Cluster of Excellence, Johannes Gutenberg-Universit\"at, D-55099 Mainz, Germany}

\date{\today}

\begin{abstract}

We propose a direct search for the X17 particle, which was conjectured to explain the ATOMKI $^8$Be and $^4$He anomalies, through the dilepton photoproduction process on a nucleon in the photon energy range below or around the pion production threshold. 
For the scenarios of either pseudoscalar, vector, or axial-vector quantum numbers of the conjectured X17, we use existing constraints to estimate the X17 signal process.
For dilepton resolutions which have been achieved in previous experiments, a signal-to-background ratio of up to an order of magnitude is found for a neutron target, and in particular for the pseudoscalar and vector X17 scenarios. 

\end{abstract}
\maketitle

A few years ago, the ATOMKI group measured electron-positron angular correlations for two magnetic dipole transitions to the ground state taking place in \textsuperscript{8}Be~\cite{Krasznahorkay:2015iga}. 
At large angles the correlation significantly deviated from the expectation for the transition from the predominantly isoscalar excited state at 18.15\,MeV to the \textsuperscript{8}Be ground state, whereas no signal was found in the decay of the predominantly isovector excited state at 17.64\,MeV. In a second experiment with an improved and independent setup, the signal for the transition from the 18.15\,MeV state was confirmed~\cite{Krasznahorkay:2018snd,Krasznahorkay:2019lgi}. 
Furthermore, the same collaboration reported an excess with around $7\sigma$ significance in a transition in~\textsuperscript{4}He, around the same $e^+e^-$ invariant mass~\cite{Krasznahorkay:2019lyl,Krasznahorkay:2021joi}. Both observations were conjectured by the authors as being due to the emission of a new boson with mass around 17\,MeV, denoted as X17. 
	
In view of a vigorous program worldwide to search for dark sector particles with a very weak coupling to Standard Model particles, from sub-eV mass scales to multi-TeV mass scales~\cite{Battaglieri:2017aum,Beacham:2019nyx,Agrawal:2021dbo}, the ATOMKI observations have sparked the prospect that the conjectured X17 might fall in this category. Based on angular momentum and parity conservation in the observed nuclear transitions, the hypothetical X17 could be a pseudoscalar ($J^P=0^-$) axion-like particle (ALP), a vector particle ($J^P=1^-$), or an axial-vector particle ($J^P=1^+$), and a variety of theoretical explanations along these lines have been proposed, see Refs.~\cite{Feng:2016jff,Feng:2016ysn,Ellwanger:2016wfe,Kozaczuk:2016nma,Alves:2017avw,Dror:2017ehi,DelleRose:2017xil,DelleRose:2018eic,DelleRose:2018pgm,Bordes:2019qjk,Nam:2019osu,Kirpichnikov:2020tcf,Feng:2020mbt,Alves:2020xhf,Fayet:2020bmb,Viviani:2021stx} among others.  
Several of these new physics explanations were challenged however, see e.g.~\cite{Zhang:2020ukq,Hayes:2021hin}, motivating to further scrutinize the energy dependence of the nuclear $(p, \gamma)$ reactions which led to the above observations.  
On the experimental side, direct searches by the NA64 Collaboration at CERN have not found any X17 evidence so far~\cite{NA64:2018lsq,NA64:2019auh}, putting constraints on the allowed parameter ranges for new physics explanations. 
Furthermore, X17 searches are part of an ongoing large scale effort at many facilities in searches for feebly-interacting particles, see \cite{Agrawal:2021dbo} for a recent review. 

In this Letter, we propose a direct search for the conjectured X17 particle through the dilepton photoproduction on a nucleon, the $\gamma N \to e^+ e^- N$ process, in the 100 - 150\,MeV photon energy range, below or around the production threshold for pions, at high-luminosity fixed target electron scattering facilities.  
For the scenarios of either pseudoscalar, vector, or axial-vector quantum numbers for the conjectured X17 in the ATOMKI \textsuperscript{8}Be anomaly, we use existing constraints to provide an estimate for the X17 signal in the $\gamma N \to e^+ e^- N$ process. For each of the three scenarios we compare this signal to the electromagnetic background for both a proton and a quasi-free neutron target, and provide an experimental outlook. 
For the pseudoscalar scenario we follow the model of \emph{Alves \& Weiner} \cite{Alves:2017avw}, for the vector case we adopt the model proposed by \emph{Feng et al.} \cite{Feng:2016jff,Feng:2016ysn}, and for the axial-vector scenario we rely on the investigation of the \textsuperscript{8}Be anomaly by \emph{Kozaczuk et al.} \cite{Kozaczuk:2016nma}.

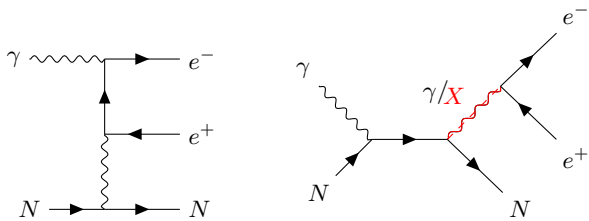
\begin{figure}
		\begin{minipage}[t]{0.49\linewidth}
    		\begin{tikzpicture}
    			\begin{feynman}[small]
    				\vertex(a){\(N\)};
    				\vertex[right=of a] (b);
    				\vertex[right=of b] (c){\(N\)};
    				\vertex[above=of b] (d);
    				\vertex[above=of d] (e);
    				\vertex[left=of e] (f){\(\gamma\)};
    				\vertex[right=of e] (g){\(e^-\)};
    				\vertex[right=of d] (h){\(e^+\)};
    				
    				\diagram* {
    					(a)-- [fermion] (b),
    					(b)-- [fermion] (c),
    					(d)-- [boson] (b),
    					(d)-- [fermion] (e),
    					(f)-- [boson] (e),
    					(e)-- [fermion] (g),
    					(d)-- [anti fermion] (h),};
    			\end{feynman}
    		\end{tikzpicture}
		\end{minipage}
		\hfill
		\begin{minipage}[t]{0.49\linewidth}
			\begin{tikzpicture}
				\begin{feynman}[small]
					\vertex(a){\(N\)};
					\vertex[above right=of a   ] (b);
					\vertex[above left=of b] (f1){\(\gamma\)};
					\vertex[right=of b] (c);
					\vertex[above right=of c] (d);
					\vertex[above right=of d] (e){\(e^-\)};
					\vertex[below right=of d] (f){\(e^+\)};
					\vertex[below right=of c] (f3){\(N\)};
					
					\diagram* {
						(a)-- [fermion] (b),
						(f1)-- [boson] (b),
						(b)-- [fermion] (c),
						(c)-- [boson, edge label=\(\gamma/\ \ \)] (d),
						(c)-- [scalar, boson, edge label=\(X\),red] (d),
						(c)-- [fermion] (f3),
						(d)-- [fermion] (e),
						(d)-- [anti fermion] (f),};
				\end{feynman}
			\end{tikzpicture}
		\end{minipage}
		\caption{Right panel: direct tree-level Feynman diagram for the signal process $\gamma N \rightarrow e^+e^- N$ via a new physics particle $X$. The process on the right panel with a photon ($\gamma$) instead of $X$ and the Bethe-Heitler process on the left panel are the main QED background processes.
		The crossed diagrams are not shown explicitely.}
		\label{fig:Feynman}
	\end{figure}

In order to estimate the possible X17 signal in the $\gamma N \to e^+ e^- N$ process, see right panel in Fig.~\ref{fig:Feynman}, we start from the reported ATOMKI value for the  ratio of the decay rate via the new boson, denoted by $X$ in the following, to the $\gamma$ decay rate of the (predominantly) isoscalar transition in $^8$Be~\cite{Krasznahorkay:2019lgi}:
\begin{eqnarray}
&&\Gamma \left(^8\text{Be}(18.15) \to \, ^8\text{Be}(\text{g.s.}) X \right)
\nonumber \\
&&\quad = \left( 6 \pm 1\right) \times 10^{-6} \; \Gamma \left(^8\text{Be}(18.15) \to \,  ^8\text{Be}(\text{g.s.}) \gamma \right), 
\nonumber \\
&&\quad = \left( 1.2 \pm 0.2 \right) \times 10^{-5} \quad \rm{eV},
\label{eq:ratio}
\end{eqnarray}
assuming a branching ratio $BR(X \rightarrow e^+e^-) = 1$, 
and using the reported value of the new 
boson mass~\cite{Krasznahorkay:2019lgi}:
	\begin{equation}
	\label{eq:mass}
	m_{X} = 17.01(16)\,\text{MeV}.
	\end{equation}
	For $BR(X \rightarrow e^+e^-) = 1$, the X17 signal in the $\gamma N \to e^+ e^- N$ process depends only on the $X$ coupling to the nucleon, but not on the value of the electron coupling. We  subsequently discuss the quark and nucleon couplings for the three possible $X$ quantum number scenarios. 
	
	Following \emph{Alves \& Weiner} \cite{Alves:2017avw} for the pseudoscalar ALP scenario, the coupling of an ALP $X$ to the nucleon isospin doublet $N$ is described by:
	\begin{equation}
		\mathcal{L}_{PS} = i g_{XNN} \bar{N} \gamma_5 N X,
	\end{equation}
	where the coupling constant can be separated in an isoscalar and an isovector part,
	\begin{equation}
		g_{XNN} = g_{XNN}^{(0)} + g_{XNN}^{(1)}\tau_3,
	\end{equation}
	with isospin-space Pauli-matrix $\tau_3$ and
	\begin{equation}
	g_{XNN}^{(1)} = \frac{m_N}{f_{\pi}}(\Delta u - \Delta d) \theta_{X \pi},
	\label{eq:psivcoupling}
		\end{equation}
	with nucleon mass $m_N$, pion decay constant $f_{\pi}\approx92.4\,$MeV, 
	axial charges $\Delta u, \Delta d$, and ALP-$\pi^0$ mixing angle $\theta_{X\pi}$. The isovector combination of axial charges is well determined from nuclear $\beta$-decay, $(\Delta u - \Delta d) \simeq 1.27$, while searches for the decay $\pi^+ \rightarrow e^+ \nu_e X \rightarrow e^+ \nu_e e^+ e^-$ by the SINDRUM Collaboration~\cite{SINDRUM:1986klz} put the very strong constraint $|\theta_{X\pi}| \lesssim (0.5-0.7) \times 10^{-4}$~\cite{Alves:2017avw}. Such a pion-phobic ALP could also explain the KTeV anomaly~\cite{KTeV:2006pwx} in another pion decay, $\pi^0 \rightarrow e^+e^-$. 
	By taking $\lvert\theta_{X\pi}\rvert \lesssim 0.5 \times 10^{-4}$, Eq.~(\ref{eq:psivcoupling}) leads to the bound:	
	\begin{equation}
		\lvert g_{XNN}^{(1)}\rvert \lesssim 0.6 \times 10^{-3}.\label{eq:gaNN1}
	\end{equation}
	The isoscalar coupling $g_{XNN}^{(0)}$ is then  constrained by the ATOMKI results. The ratio of ALP to M1 photon emission rates with isospin change $\Delta T = 0, 1$ was calculated following the early work of \emph{Donnelly et al.}~\cite{Donnelly:1978ty}:
	\begin{equation}
	\label{eq:ratioDT}
		\frac{\Gamma_X}{\Gamma_{\gamma}}\bigg\vert_{\Delta T} = \frac{1}{2\pi \alpha} \left(\frac{g_{XNN}^{(\Delta T)}}{\mu^{(\Delta T)}-\eta^{(\Delta T)}}\right)^2 
		\biggl[1-\Bigl(\frac{m_X}{\Delta E}\Bigr)^2\biggr]^{3/2}, 
	\end{equation}
where $\alpha \approx 1/137$ is the fine-structure constant, and $\Delta E$ is the excitation energy of the corresponding nuclear level. Furthermore in Eq.~(\ref{eq:ratioDT}), the parameters $\mu$ and $\eta$ are the form factor values at momentum transfer $\sim \mathcal{O}(17\,\text{MeV})^2 \approx 0$ and are related to nuclear magnetic moments and the ratio of convection to magnetization currents, respectively. They have been estimated as $\mu^{(0)} = 0.88$, $\mu^{(1)} = 4.7$, $\eta^{(0)} = 1/2$, while $\eta^{(1)}$ can be neglected compared to $\mu^{(1)}$ as a first estimate~\cite{Donnelly:1978ty}.
	
Due to isospin mixing of the $^8$Be excited states at 18.15\,MeV (predominantly isoscalar) and 17.64\,MeV (predominantly isovector), the comparison with the measured decay rates involves an isospin mixing angle $\theta_{1^+}$,  which we take as $\sin \theta_{1^+} = 0.35 (8)$, following the analysis of~\cite{Kozaczuk:2016nma}. The ATOMKI value for  the transition ratio of the 18.15\,MeV state, given in Eq.~(\ref{eq:ratio}), then yields for the isoscalar coupling $g_{XNN}^{(0)}$ the range shown in Table~\ref{tab:couplingvalues}. Furthermore, the value of the transition ratio for the 17.64\,MeV state in $^8$Be is found to be: 
	\begin{equation}
		\frac{\Gamma_X}{\Gamma_{\gamma}}\bigg\vert_{{^8\text{Be}}(17.64)}\approx (0.4-9.3) \times 10^{-8},
	\end{equation}
which is one to two orders of magnitude smaller than the one for the 18.15\,MeV state, and consistent with the ATOMKI non-observation of an $X$ particle in the 17.64\,MeV $\to$ g.s. transition in $^8$Be.

In a later work~\cite{Alves:2020xhf}, \emph{Alves} used a similar estimate from Ref.~\cite{Donnelly:1978ty} for the $0^-\rightarrow 0^+$ transition in \textsuperscript{4}He and found that the model can consistently explain both anomalies.

We next discuss the vector scenario for the $X$ particle, proposed in Refs.~\cite{Feng:2016jff,Feng:2016ysn}, in which the coupling to quarks is given by:
	\begin{equation}
		\mathcal{L}_V = - e X_{\mu} \sum_q \varepsilon_q  \bar{q}\gamma^{\mu} q.
	\end{equation}
The nucleon couplings are then obtained from the quark couplings as: $ \varepsilon_p= 2\varepsilon_u + \varepsilon_d$ and $\varepsilon_n = \varepsilon_u + 2\varepsilon_d$. 
In the limit of no isospin mixing or breaking, the nuclear part of the matrix element in the decay rate ratio $\Gamma_X/\Gamma_{\gamma}$ cancels out, simply yielding for the isoscalar state:
\begin{eqnarray}
		\frac{\Gamma_X}{\Gamma_{\gamma}}\bigg\vert_{^8\text{Be}(18.15)} 
		&=& (\varepsilon_p + \varepsilon_n)^2 \biggl[1-\Bigl(\frac{m_X}{18.15\, \mathrm{MeV}}\Bigr)^2\biggr]^{3/2}, \quad \quad
\end{eqnarray}
which constrains the sum $(\varepsilon_p + \varepsilon_n)$. 
The expression becomes slightly more complicated when including isospin mixing and isospin breaking. In our numerical analysis we follow Ref.~\cite{Feng:2016ysn}, using their  breaking parameter $\kappa=0.549$, and the above mentioned isospin mixing parameter $\theta_{1+}$. The inclusion of the constraint provided by the NA48/2 experiment, which looked for the neutral pion decay $\pi^0 \rightarrow \gamma (X\rightarrow e^+e^-)$ \cite{NA482:2015wmo}, leads to the protophobia condition ($\varepsilon_p \ll \varepsilon_n$)~\cite{Feng:2016ysn}:
	\begin{equation}
		\lvert\varepsilon_p\rvert \lesssim 1.2\times 10^{-3}.
	\end{equation}
	The ATOMKI decay rate for the 18.15\,MeV transition then provides a lower limit on the $X$ coupling to the neutron $\varepsilon_n$, as given in Table~\ref{tab:couplingvalues}.

	Thirdly, we also discuss the scenario when the $X$ boson has purely axial-vector interaction with quarks:
	\begin{equation}
	\mathcal{L}_A = -X_{\mu}\sum_{q} g_q \bar{q} \gamma^{\mu} \gamma_5 q.
	\end{equation}

	\emph{Kozaczuk et al.} calulated the decay widths of the $1^+$  \textsuperscript{8}Be states to the $0^+$ g.s. in  \textsuperscript{8}Be via such an axial-vector boson to be~\cite{Kozaczuk:2016nma}: 
	\begin{equation}
	\Gamma_X = \frac{\lvert k_X \rvert}{18 \pi} \left[2 + \left(\frac{ \Delta E}{m_X}\right)^2 \right] \Big| a_n \left\langle 0\lVert\sigma^n\rVert 1\right\rangle + a_p \left\langle 0\lVert\sigma^p\rVert 1\right\rangle \Big|^2,
	\end{equation}
	with $\lvert k_X \rvert = \Delta E \, [1 - (m_X/\Delta E)^2]^{1/2}$, and 
	where the nucleon couplings are expressed as:
	\begin{equation}
	a_{p,n} = \sum_{q=u,d,s}\Delta q^{(p,n)}g_q. \label{eq:Deltaq}	\end{equation}
	As the axial-vector interaction couples to the spin, the light quark couplings are weighted by the axial charges $\Delta q$, for which the recent results from~\cite{Bishara:2016hek} are used:
	\begin{align}
		\label{eq:nucleon coefficients}
		\Delta u^{(p)} =&\ \Delta d^{(n)} =  0.897 (27),\nonumber\\
		\Delta d^{(p)} =&\ \Delta u^{(n)} = -0.367 (27),
	\end{align}
	and where the small $\Delta s^{(p)}$ is neglected for simplicity. The reduced nuclear matrix elements $\left\langle 0\lVert\sigma^{p,n}\rVert 1\right\rangle$ were estimated by \emph{Kozaczuk et al.}, using the isospin mixing value $\sin \theta_{1^+} = 0.35$. For the two states at 18.15 and 17.64\,MeV, they were obtained from fig.~2 in \emph{Kozaczuk et al.} \cite{Kozaczuk:2016nma} to be
	\begin{eqnarray}
		\label{eq:matrix elements}
		\left\langle 0\lVert\sigma^p\rVert1 (17.64)\right\rangle &=&\ 0.100(18),\nonumber \\  
		\left\langle 0\lVert\sigma^n\rVert 1 (17.64)\right\rangle &=&  -0.070(11), \nonumber\\
		\left\langle 0\lVert\sigma^p\rVert 1 (18.15)\right\rangle &=&\ -0.044(13), \nonumber \\ 
		\left\langle 0\lVert\sigma^n\rVert 1 (18.15)\right\rangle &=&  -0.130(21).
	\end{eqnarray}
	Similar to the vector scenario above, the \textsuperscript{8}Be ATOMKI experiment constrains roughly the (isoscalar) sum of the nucleon couplings. 
	Assuming $a_p = a_n$ in this work (corresponding with $g_u = g_d$), we then derive a bound on its value from the observed decay rates from the ATOMKI experiment for the 18.15\,MeV and 17.64\,MeV states in $^8 Be$, and the corresponding values for $a_{p, n}$ and $g_{u, d}$ are shown in Table~\ref{tab:couplingvalues}.

\begin{table}
		\caption{The values for the $X$ coupling constants to the nucleon in the three discussed scenarios of $J^P_X$ quantum numbers. The left column shows the couplings using the central value for the $X$ mass, the right column the values using a $1 \sigma$ variation on the $m_X$ value.}
		\begin{tabular}{c|c|c}
			\hline
			\hline
			$J_X^{P}$ & $m_{X}=17.01\,$MeV & $1\sigma$ uncertainty in $m_X$\\
			\hline
			\hline
			$0^-$ & \multicolumn{2}{c}{$|g_{XNN}^{(1)}|=(0-0.6)\times 10^{-3}$}\\
			 & $g_{XNN}^{(0)}=(3.0-4.0)\times 10^{-3}$ & $g_{XNN}^{(0)}=(2.7-4.4)\times 10^{-3}$\\
			 \hline
			$1^-$ & \multicolumn{2}{c}{$|\varepsilon_p|=(0-0.12)\times 10^{-2}$}\\
			 & $|\varepsilon_n|=(1.2-1.7)\times 10^{-2}$ & $|\varepsilon_n|=(1.1-1.9)\times 10^{-2}$\\
			 \hline
			$1^+$ & $g_{u,d}=(3.9-10.1)\times 10^{-5}$ & $g_{u,d}=(3.7-10.5)\times 10^{-5}$\\
			 & $a_{p,n}=(1.9-5.9)\times 10^{-5}$ & $a_{p,n}=(1.8-6.1)\times 10^{-5}$\\
			\hline
			\hline
		\end{tabular}
		\label{tab:couplingvalues}
\end{table}
	
Within the three discussed scenarios, we next estimate the signal for the photoproduction of a X17 particle in the
$\gamma N \rightarrow e^+e^-N$ reaction, where $N$ is either a proton or neutron, as shown in Fig.~\ref{fig:Feynman} (right panel). For photon energies below and around pion production threshold, the Feynman diagrams for the two leading background processes are also shown in Fig.~\ref{fig:Feynman}: the Bethe-Heitler process (left) as well as the Born process (right), which has the same topology as the signal process, with the $X$ particle replaced by a photon. In both cases there is also a crossed diagram, which is not shown in Fig.~\ref{fig:Feynman}.

\begin{figure}
		\centering
		\includegraphics[width=.45\textwidth]{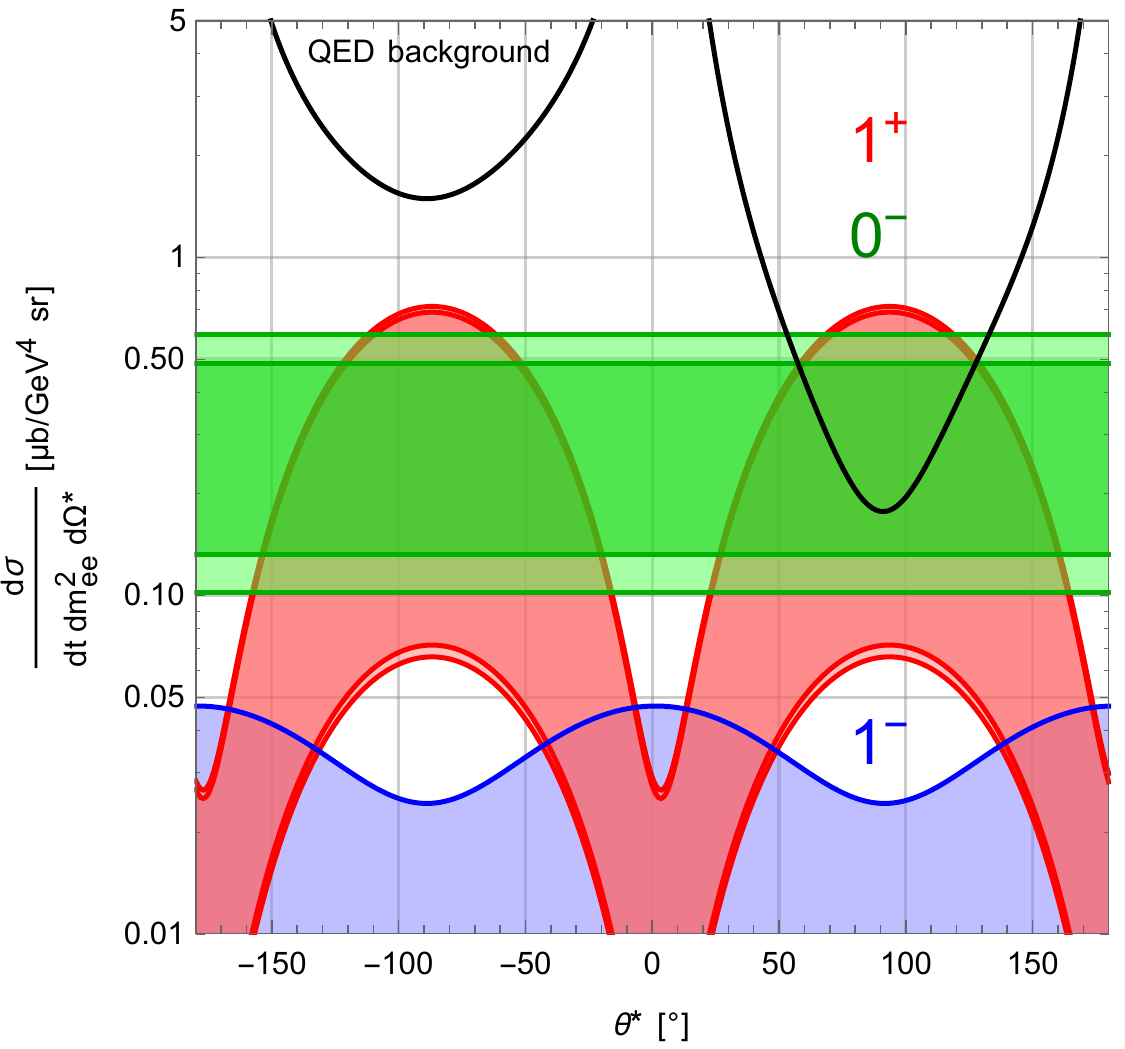}
		\includegraphics[width=.45\textwidth]{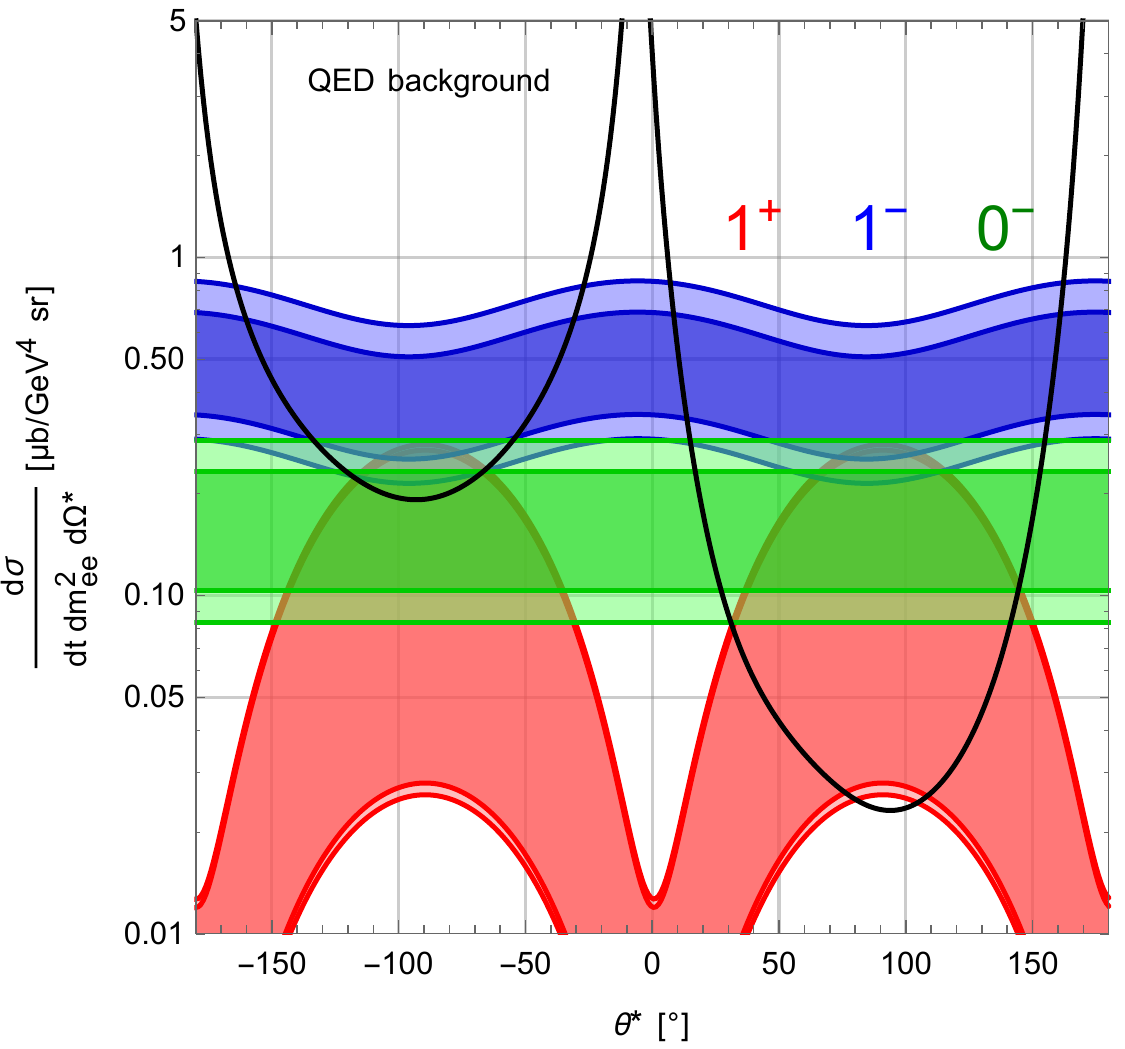}
		\caption{
		Cross section estimate of the X17 angular distribution in the $\gamma N \to e^-e^+ N$  reaction for photon energy $E_\gamma = 0.15$\,GeV and momentum-transfer to nucleon $-t = 0.04$\,GeV$^2$,  
		with $N=p(n)$ at $\phi^*=45^{\circ}(180^{\circ})$ in the upper (lower) panels. 
		The signal cross section is averaged over a bin of 0.2\,MeV around $m_{ee} = m_X$ and is shown for three X17 scenarios: as a pseudoscalar ALP (green), as a vector state (blue), and as an axial-vector state (red), together with the QED background (black curve). The inner (outer) bands depict the uncertainty due to the \textsuperscript{8}Be decay width (X17 mass).
		}
		\label{fig:angdistp}
	\end{figure}
	
The differential cross section for this process is given by:
	\begin{equation}
		\frac{d\sigma}{dt dm_{ee}^2 d\Omega^*} = \frac{1}{64} \frac{1}{(2\pi)^4} \frac{1}{(2m_N E_{\gamma})^2} \overline{|\mathcal{M}|^2}\label{eq:yN-eeN_cross},
	\end{equation}
	with $E_\gamma$ the lab energy of the the incoming photon, $t$ the four-momentum transfer to the nucleon and $m_{ee}$ the invariant mass of the dilepton pair. $\theta^*$ is the electron polar angle in the $e^+e^-$ rest frame with respect to the lab momentum direction of the $e^+e^-$ pair, and $\phi^*$ is the azimuth angle of the $e^+ e^-$ decay plane w.r.t. the plane of the incoming photon and the lab direction of the dilepton pair momentum. Furthermore, $\overline{|\mathcal{M}|^2}$ is the squared matrix element averaged over initial and summed over final spins. 
	
	Fig.~\ref{fig:angdistp} shows the angular dependence of the differential cross section for $E_{\gamma}=0.15\,$GeV and $-t=0.04\,$GeV$^2$ with angles $\phi^*=45^{\circ}(180^{\circ})$ that allow maximizing the signal-to-background-ratio for a proton (neutron) target. 
	The $X$ signal cross section is averaged over a bin $\Delta m_{ee} = 0.2$\,MeV around $m_{ee} = m_X$, as such energy resolution in the $e^+e^-$-invariant mass was already achieved at a dark photon search experiment at MAMI~\cite{Merkel:2014avp}. 
	The QED background process is depicted in black, while the signal process is shown for the pseudoscalar (green), vector (blue) and axial-vector (red) scenarios. 
	Both plots in Fig.~\ref{fig:angdistp} show two consecutive error bands graded by color. The darkest inner bands correspond with the couplings in the left column of Table~\ref{tab:couplingvalues}, which were obtained by fixing $m_{X}=17.01\,$MeV and varying over the range of the ATOMKI $^8$Be decay rate of Eq.~\eqref{eq:ratio}. The outer bands were obtained by also considering a $1 \sigma$ variation of the mass $m_{X}$ according to Eq.~\eqref{eq:mass}, corresponding to the couplings in the right column of Table~\ref{tab:couplingvalues}. The correlation between the assumed mass and the best-fit decay rate was neglected here.
	One notices from Fig.~\ref{fig:angdistp}, that 
	for a proton target and for $\Delta m_{ee} = 0.2$\,MeV, the signal is at best of the order of the background for the pseudoscalar or axial-vector X17 scenario around $\theta^*=90^{\circ}$, while the signal of a protophobic vector boson cannot be expected to be measurable. For a neutron target, however, the background is considerably smaller in the same kinematical configuration due to the absence of a charge coupling to the neutron, and all three X17-scenarios could leave a significant signal, with the vector scenario nearly an order of magnitude above the background around $\theta^*=90^{\circ}$. The different angular dependencies for the three scenarios would allow to determine the quantum numbers of an $X$ particle thus produced. 
One can optimize such a search experiment, and further increase the signal-to-background ratio by a better energy resolution. For $\Delta m_{ee} = 0.1$\,MeV e.g., one would increase the signal-to-background ratio by a factor of 2 compared to Fig.~\ref{fig:angdistp}.  

In conclusion, we proposed a direct search for the X17 particle, which was conjectured to explain the ATOMKI $^8$Be and $^4$He anomalies, 
through the dilepton photoproduction on a nucleon in the photon energy range below or around the pion production threshold. 
We analyzed the discovery potential for three $J^P_X$ quantum number scenarios of an X17. For the cases of a pseudoscalar, vector, and axial-vector, we calculated the signal process by estimating the coupling constants from the observed $^8$Be decay rate value. 
The $\gamma N \to e^+e^- N$ signal cross section was compared to the expected background for a dilepton mass resolution which has been achieved before. A  discovery potential was found when considering the process for a neutron target, and in particular for the pseudoscalar and vector scenarios for an X17. Such a search experiment can be performed at electron accelerators such as e.g. MAMI or MESA.
\\
\\
This work was supported by the Deutsche Forschungsgemeinschaft (DFG, German Research Foundation), in part through the Collaborative Research Center [The Low-Energy Frontier of the Standard Model, Projektnummer 204404729 - SFB 1044], and in part through the Cluster of Excellence [Precision Physics, Fundamental Interactions, and Structure of Matter] (PRISMA$^+$ EXC 2118/1) within the German Excellence Strategy (Project ID 39083149).

\end{document}